\documentclass[%
 reprint, showkeys,
superscriptaddress,
 amsmath,amssymb,
 aps,
floatfix,
]{revtex4-2}
\usepackage{gensymb}
\usepackage{setspace}
\usepackage{dirtytalk}
\usepackage{natbib}
\usepackage{amsmath}
\usepackage{mathtools}
\usepackage{upgreek}
\usepackage[pdftex]{graphicx}
\usepackage{float}
 \usepackage[left]{lineno}
\usepackage{graphicx}
\usepackage{dcolumn}
\usepackage{bm}
\usepackage[colorlinks=true, urlcolor=blue, linkcolor=blue, citecolor=blue]{hyperref}
\hypersetup{%
    bookmarksopen=false,
    bookmarksnumbered=true,
    pdfnewwindow=true,
    unicode=false,
    colorlinks=true,%
    citecolor=blue,
    linkcolor=black,
    urlcolor=blue,
    filecolor=blue
    }
\usepackage{xcolor}%
\usepackage{soul}

\definecolor{Blue}{rgb}{0.3,0.3,0.9}
\definecolor{Red}{rgb}{0.9,0.3,0.3}
\definecolor{Green}{rgb}{0.3,0.6,0.3}

\begin{document}

\preprint{APS/123-QED}

\title{Quantum Criticality in Monolayer Amorphous Carbon  
}

\author{Rejaul SK}
\affiliation{%
Division of Physics and Applied Physics, School of Physical and Mathematical Sciences, Nanyang Technological University, Singapore 637371, Singapore
}%
\author{Hanning Zhang}
\affiliation{Institute of PhysicsEcole, Polytechnique Fédérale de Lausanne (EPFL), Lausanne CH-1015, Switzerland}%
\author{ Artem K. Grebenko}
\affiliation{%
Department of Physics, National University of Singapore, Singapore 117551, Singapore
}%
\author{Arsen Herasymchuk}
\affiliation{Department of Physics, University of Z\"{u}rich, CH-8057 Z\"{u}rich, Switzerland
}%
\author{Ranjith Shivajirao}
\affiliation{%
Division of Physics and Applied Physics, School of Physical and Mathematical Sciences, Nanyang Technological University, Singapore 637371, Singapore
}%
\author{Hongji Zhang}
\affiliation{Department of Materials Science and Engineering, National University of Singapore, Singapore
}%
\author{Abee Nelson}
\affiliation{%
Division of Physics and Applied Physics, School of Physical and Mathematical Sciences, Nanyang Technological University, Singapore 637371, Singapore
}%
\author{Zheng Jue Tong}
\affiliation{%
Division of Physics and Applied Physics, School of Physical and Mathematical Sciences, Nanyang Technological University, Singapore 637371, Singapore
}
\author{Gagandeep Singh}
\affiliation{%
Division of Physics and Applied Physics, School of Physical and Mathematical Sciences, Nanyang Technological University, Singapore 637371, Singapore
}
\author{Naoto Kimiuchi}
\affiliation{Sanken, Osaka University, Ibaraki, Japan
}%
\author{Yuta Sato}
\affiliation{Nanomaterials Research Institute, National Institute of Advanced Industrial Science and Technology (AIST), Tsukuba, Japan
}%
\author{Kazutomo Suenaga}
\affiliation{Sanken, Osaka University, Ibaraki, Japan
}%
\author{Chee Tat Toh}
\affiliation{Department of Materials Science and Engineering, National University of Singapore, Singapore
}%

\author{Rudolf A. R\"{o}mer}
\affiliation{%
 Department of Physics, University of Warwick, Gibbet Hill Road, Coventry, CV4 7AL, UK
}%
\author{Shaffique Adam}
\affiliation{Department of Physics, Washington University in St. Louis, St. Louis, Missouri 63130, USA
}%
\author{Oleg V Yazyev}
\affiliation{Institute of PhysicsEcole, Polytechnique Fédérale de Lausanne (EPFL), Lausanne CH-1015, Switzerland}%
\author{ Barbaros \"{O}zyilmaz}
\email{barbaros@nus.edu.sg}
\affiliation{Department of Physics, National University of Singapore, Singapore 117551, Singapore
}%
\author{Bent Weber}%
\email{b.weber@ntu.edu.sg}
\affiliation{%
Division of Physics and Applied Physics, School of Physical and Mathematical Sciences, Nanyang Technological University, Singapore 637371, Singapore
}%

\date{\today}

\begin{abstract}

Amorphous solids represent the extreme limit of broken translational symmetry, in which the absence of long-range order removes well-defined crystal momenta and invalidates the Bloch description of electronic states. Monolayer amorphous carbon (MAC) has emerged as a unique realization of a strictly two-dimensional (2D) amorphous lattice defined by a structurally contiguous but topologically disordered $sp^2$-bonded random network devoid of any defined long-range crystal symmetry. From atomic-resolution measurements of multifractal wavefunctions, we show that disorder in MAC effectively localizes the low-energy part of the electronic spectrum but retains an extended critical-like state near the band centre ($E\sim 0$). We conjecture that this state is protected from topological disorder by remnant chiral symmetry surviving within the continuous random network, described by a Wess-Zumino-Witten (WZW) topological term. Near criticality, we verify the multifractal scaling relation $\eta = -\Delta_2$, providing quantitative agreement between independently measured spatial correlation decay and multifractal scaling exponents. Our results are confirmed by atomistic tight-binding calculations that closely mirror the multifractal scaling near $E\sim 0$. Our results establish MAC as the first strictly 2D amorphous electronic system to exhibit Anderson criticality driven purely by topological disorder

\end{abstract}

\keywords{Monolayer amorphous carbon, Topological disorder,  Anderson localization, Wavefunction criticality 
}
\maketitle


\section{Introduction}
Anderson localization \textendash\ the spatial localization of wave packets by scattering \textendash\ is as ubiquitous as quantum mechanical waves themselves~\cite{Krameri1993}. First predicted in 1958 for electron waves in random lattices~\cite{anderson_absence_1958}, 
its universality has been demonstrated across a wide range of platforms~\cite{abrahams_scaling_1979, evers_anderson_2008}. Examples include ultracold atomic gases in disordered optical potentials~\cite{billy_direct_2008, roati_anderson_2008,deissler_delocalization_2010}, photonic crystals and optical waveguide arrays~\cite{segev_anderson_2013,khanikaev_photonic_2013,sapienza_cavity_2010,schwartz_transport_2007}, acoustic and elastic media~\cite{hu_localization_2008}, and more recently, superconducting~\cite{malavi_enhancement_2025,zhao_disorder-induced_2019} and topological systems~\cite{li_topological_2009,meier_observation_2018,zhang_experimental_2021}. Since dimensionality and quantum interference are intimately connected, they play a decisive role in Anderson localization, determining whether disorder localizes electronic wavefunctions or whether extended states can prevail~\cite{abrahams_scaling_1979}.

In 2D, quantum interference drives all electronic states towards localization in the presence of arbitrarily weak generic disorder~\cite{abrahams_scaling_1979}. Modern theoretical approaches have proposed that the presence or absence of specific symmetries fundamentally determines the fate of disordered electronic states in 2D. The use of nonlinear sigma models has led to a classification of disordered systems into distinct symmetry classes, such as orthogonal, unitary, and symplectic, based on time-reversal, spin-rotation, and chiral symmetries~\cite{evers_anderson_2008,altland_nonstandard_1997} that can protect electronic states from localization. A  notable example is 2D graphene, in which chiral symmetry of the low-energy spectrum due to the bipartite lattice suppresses backscattering of Dirac fermions, favouring weak anti-localization under smooth disorder~\cite{ryu_mathbbz_2_2007,punnoose_metal-insulator_2005, das_sarma_electronic_2011, garcia_adatoms_2014,fan_anderson_2014,xiong_anderson_2007,garcia_adatoms_2014,Gonzalez-Santander2013LocalisationFlakes}.
However, this foundational result, relies on the assumptions that disorder manifests as random on-site potentials in an otherwise crystalline lattice with well-defined symmetries.

In amorphous materials~\cite{street1991hydrogenated,liu2014hydrogen,yaglioglu2006high, jankousky_effective_2026,corbae_observation_2023, corbae_amorphous_2023} these symmetries are maximally broken as lattice connectivity itself is disordered~\cite{van2012insulating, barkema2000high}. As a result, localization emerges from topology-induced scattering, correlated structural disorder, and emergent quantum interference, and localization behavior that can differ qualitatively from the conventional Anderson picture~\cite{evers_anderson_2008,van2012insulating,Grimm1998ElectronicSystems,bhattacharjee2025anderson}.

Recently, monolayer amorphous carbon (MAC)~\cite{toh_synthesis_2020,bai_nitrogen-doped_2024} \textendash\ a graphene derivative \textendash\ has emerged as a noncrystalline yet structurally contiguous 2D random network (CRN), of $sp^2$-bonded carbon. The random distribution of lattice connectivity without definite translational and sublattice symmetry constitutes a unique form of topological disorder, defined by bond geometry rather than by random onsite potentials, as all carbon atoms remain chemically identical. Scanning tunnelling microscopy and spectroscopy (STM/STS) is a unique probe for mapping the electronic wavefunction's amplitude and spatial structure. It has been applied to probe the multifractal electronic states near the criticality in 3D ~\cite{richardella_visualizing_2010, jack_visualizing_2021} and 2D ~\cite{shin_structural-disorder-driven_2023} disordered systems. Here, we apply the same technique to probe the criticality in a strictly 2D amorphous material.

%
%

For the first time, here we tie electronic wave function structure and response to microscopically resolved topological lattice disorder of MAC as a truly 2D amorphous monolayer. Using STM and STS, we observe multifractal scaling~\cite{richardella_visualizing_2010, shin_structural-disorder-driven_2023, jack_visualizing_2021} of the wave function's probability density, a hallmark for localization near the Anderson transition \cite{mirlin_distribution_1994,Rodriguez2009a,Rodriguez2010,Rodriguez2011}. Notably, we observe critical-like extended states at the band centre ($E \sim 0$) \textendash\ a strikingly unexpected result. 
We attribute this behavior to a remnant chiral structure at low energy, enabling an effective chiral field theory with a Wess-Zumino-Witten (WZW) topological term~\cite{wess1971consequences,witten1983global}, stabilizes critical-like extended states against localization. 

Atomistic tight-binding calculations of the amorphous network corroborate our observations, revealing that connectivity-driven topological disorder governs electronic localization while remnant chiral structure provides the conditions necessary  for persistence of extended states near the band centre. Our results indicate that lattice topology and emergent Hamiltonian symmetry can play a decisive role in shaping quantum interference in 2D, providing a route to topologically protected extended-like electronic states beyond the conventional Anderson framework.


\begin{figure*}
\centering
\includegraphics[width=\textwidth]{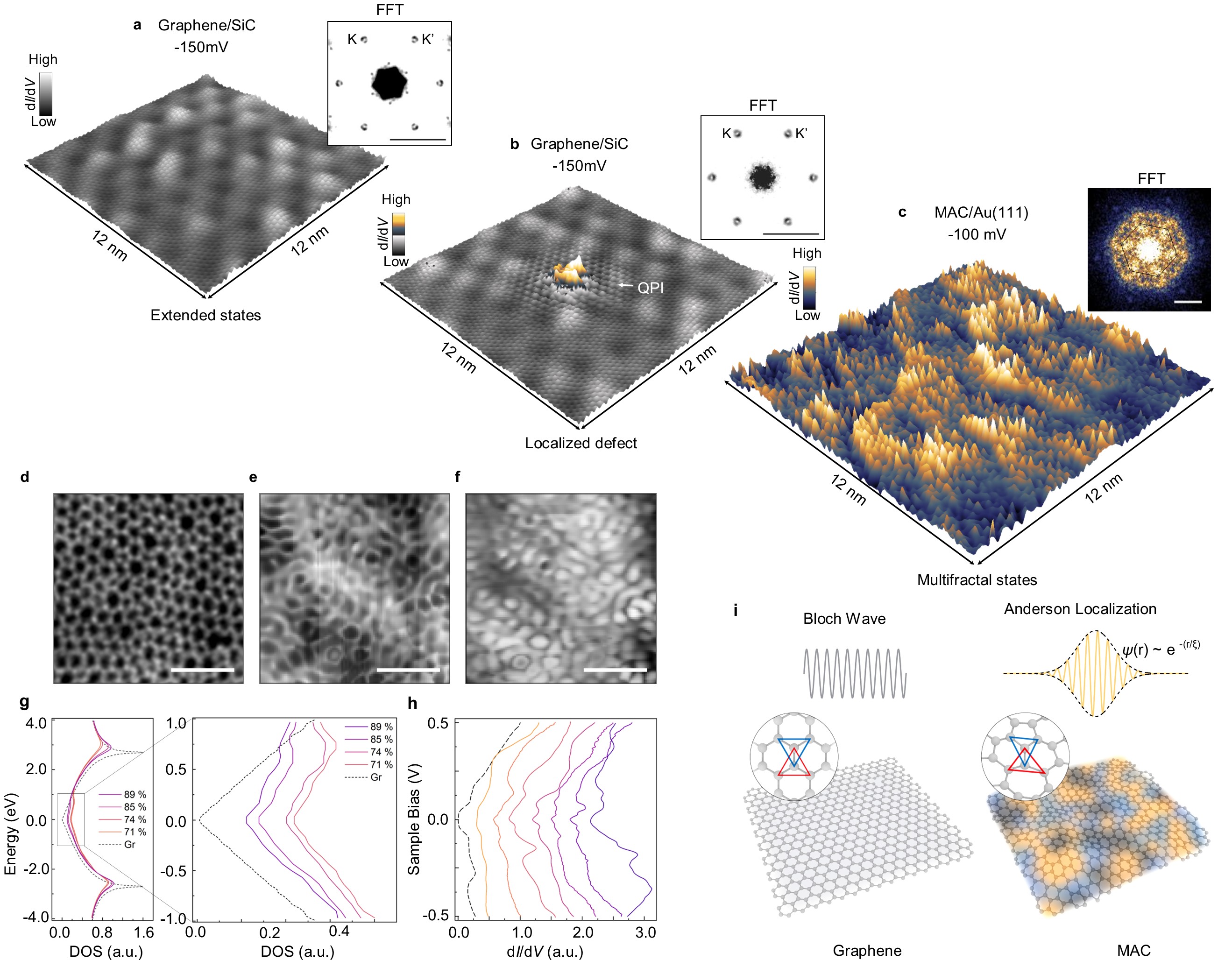}
\caption{\textbf{Microscopic signatures of critical wavefunctions in monolayer amorphous carbon:}  \textbf{a,b,} Constant current d$I$/d$V$maps on graphene on SiC, comparing a defect-free area (\textbf{a}) and a region with a point defect (\textbf{b}), give rise to quasiparticle interference (QPI). \textbf{c,} A constant current d$I$/d$V$ map on MAC, showing significant spatial modulations in the LDOS amplitude, indicative of a critical-like fluctuation. The insets show the corresponding FFTs (Scale bar = $3$ nm$^{-1}$. \textbf{d,} High-resolution Transmission electron microscope (HR-TEM) image of MAC, highlighting the topological disorder , defined by random distribution of bond length and bond angle of non-hexagonal rings. \textbf{e,f,} STM topographies of MAC (constant - height mode) at -100mV and +100mV, respectively (Scale bar = $1$ nm). \textbf{g,h} Calculated DOS of MAC, \textbf{h,} compared to measured differential conductance (dI/dV) spectra of epitaxial graphene (black dotted curve in \textbf{h,} ) and various location  on MAC (colored).\textbf{i,} Schematic representation of electron localization in MAC. }
\label{fig1}
\end{figure*}
\section{Results}

Structural characterization of MAC is shown in Fig.\ \ref{fig1}d-f. High-resolution TEM (Fig.\ \ref{fig1}d) directly confirms a continuous $sp^2$-bonded carbon network composed of threefold-coordinated atoms arranged in a distribution of $ 5 - 8$-membered rings. These non-hexagonal motifs break the sublattice symmetry locally and induce distortion in the residual six-member rings, which are otherwise uniform in the crystalline honeycomb lattice.
Constant-height STM topographic images (Fig.\ \ref{fig1}e-f), confirm the nature of the disordered ring motifs and validate the structural continuity of  monolayer amorphous carbon~\cite{toh_synthesis_2020}.
%
%
Fig.\ \ref{fig1}h compares STS point spectra at different positions on MAC (colored curves) with those in graphene (black dashed). These spectra confirm a metallic low-energy density of states without any clear signature of an energy gap. 
Different from the uniform spectrum of graphene, however, MAC exhibits strong spatial fluctuations in the local density of state (LDOS) with irregular spectral features, consistent with the strongly disordered electronic structure. The LDOS near $E_F$ is somewhat suppressed in some of the spectra, likely indicating a suppression of tunnelling into the 2D layer with finite in-plane momentum~\cite{zhang_giant_2008}.
Complementary DFT calculations (Fig.\ \ref{fig1}g) (see Methods) reproduce the experimentally observed broadened and asymmetric density of states, confirming that topological disorder dominates the electronic structure.

The strong spatial inhomogeneity of the electronic structure in MAC is further confirmed in Fig.\ \ref{fig1}a-c, where we compare LDOS maps measured on MAC with those measured on high-quality epitaxial graphene (see Methods). In contrast to the uniform LDOS maps of epitaxial graphene (Fig.\ \ref{fig1}a), the LDOS map of MAC (Fig.\ \ref{fig1}c) reveals strong spatial modulations across the surface largely without obvious periodicity, symmetry, or long-range order. This is further confirmed by the corresponding Fast Fourier transforms (FFT) (inset), not showing any distinctive features, different from those of graphene, in which quasiparticle interference (QPI)~\cite{brihuega_quasiparticle_2008} gives rise to period LDOS modulation in real-space, appearing as characteristic ring and hexagon-shaped patterns in reciprocal space due to coherent elastic scattering of delocalized Dirac electrons.

In MAC, strong spatial modulations in the LDOS arise from disorder-induced phase-interference of extended-like electronic states (similar to Fig.\ \ref{fig1}b) promoting carrier localization~\cite{das_sarma_electronic_2011}. 
\begin{figure*}
\centering
\includegraphics[width=\textwidth]{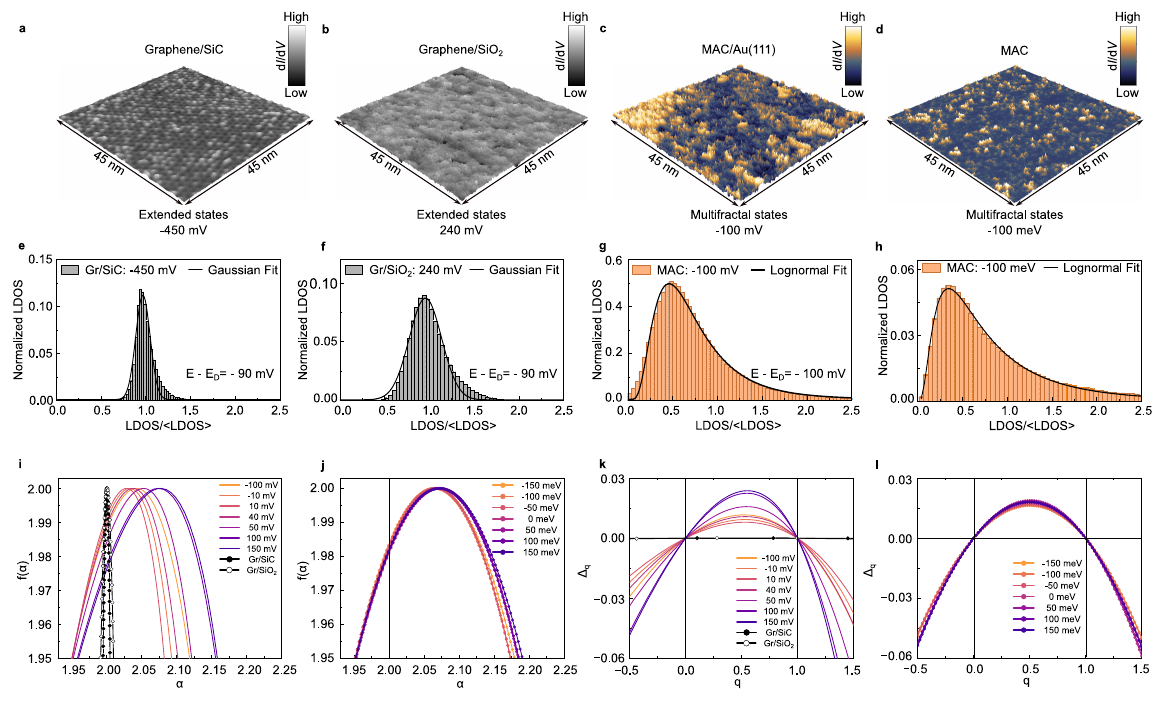}
\caption{\textbf{Fluctuation and multifractal scaling of the wavefunction probability density.} \textbf{a-d,} Measured d$I$/d$V$ maps of Gr/SiC (\textbf{a}), Gr/SiO$_2$ (\textbf{b}) and, MAC (\textbf{c}) compared to the theoretically calculated LDOS (\textbf{d}). While graphene shows a homogeneous LDOS distribution, MAC exhibits strong spatial fluctuations. \textbf{e-h,} This is further illustrated by histograms of normalized LDOS, showing Gaussian distributions (\textbf{e,f}) in graphene with a mean at 1.0, whereas MAC exhibits a lognormal distribution, indicative of critical‑like electronic fluctuations and incipient localization.\textbf{i,j,}  Measured (\textbf{i}) and calculated (\textbf{j}) singularity spectra obtained from the measured and numerically calculated LDOS maps shown in \textbf{a-d,}. MAC shows broad, asymmetric spectra characteristic of multifractal states, whereas narrow spectra in graphene confirm a 2D metallic state ($d = 2$). \textbf{k,l,} Comparison of experimental (\textbf{k}) and numerically (\textbf{l}) calculated anomalous scaling dimension \(\Delta_q\) for MAC confirming an inverted parabolic dependence on $q$, as expected in a 2D disordered system near criticality.
}
\label{fig2}
\end{figure*}
LDOS modulations as observed reflect fluctuations in the wave function amplitude, expected in electronic systems near an Anderson transition~\cite{evers_anderson_2008}. Such transition usually quantified through a statistical treatment of LDOS~\cite{rodriguez_multifractal_2011} as summarized in Fig.\ \ref{fig2}.
In metallic systems with fully extended Bloch states, the LDOS is uniformly distributed, well-approximated by a Gaussian normal distribution~\cite{USKI2002,Rodriguez2009a,terletska_systematic_2018}. 
However, near a localization transition, wavefunctions develop a \textit{multifractal} character \cite{mirlin_distribution_1994,Rodriguez2009a,Rodriguez2010,Rodriguez2011}, that is, electronic states are neither fully localized nor fully extended, but instead exhibit scale-invariant spatial fluctuations without a single characteristic length scale. As a result, the LDOS distribution becomes \textit{log-normal}, characterized by a long tail extending toward higher LDOS values and a peak biased towards zero~\cite{mirlin_distribution_1994, schubert_distribution_2010}. Such a transition is clearly observed in Fig.\ \ref{fig2}a-h where we compare LDOS distributions measured near charge neutrality in graphene and MAC, providing  evidence of localization of electronic states in MAC in excellent agreement with calculated LDOS distributions (Fig.\ \ref{fig2}d). 

A deeper understanding of the observed LDOS fluctuations can be achieved through multifractal analysis, which quantifies spatial scaling of the wavefunctions' probability density with system size \(L\) in $d$ dimensions (see supplementary information). This can be quantified by the singularity spectrum $f(\alpha)$ \textendash\ a function that expresses how the local intensity of the wavefunction scales across space in systems with complex structure near a localization \textendash\ delocalization transition. It does so by measuring the fractal dimension of the set of points that share a particular scaling exponent $\alpha$. A narrow $f(\alpha)$ (ideally a $\delta$-function) peaked at the dimension of the system ($d=2$ for graphene) describe a metallic (uniform) LDOS distribution with extended states. A broader $f(\alpha)$, on the other hand, indicates that the wavefunction has a wide distribution of local intensities, some regions strongly localized, others more extended, reflecting multifractal behavior~\cite{evers_anderson_2008}. This has been considered as sufficient experimental proof for multifractal states near criticality~\cite{richardella_visualizing_2010} both in 3D~\cite{richardella_visualizing_2010, jack_visualizing_2021} and  2D~\cite{shin_structural-disorder-driven_2023} electronic systems.
%

\begin{figure*}
\centering
\includegraphics[width=1.0\textwidth]{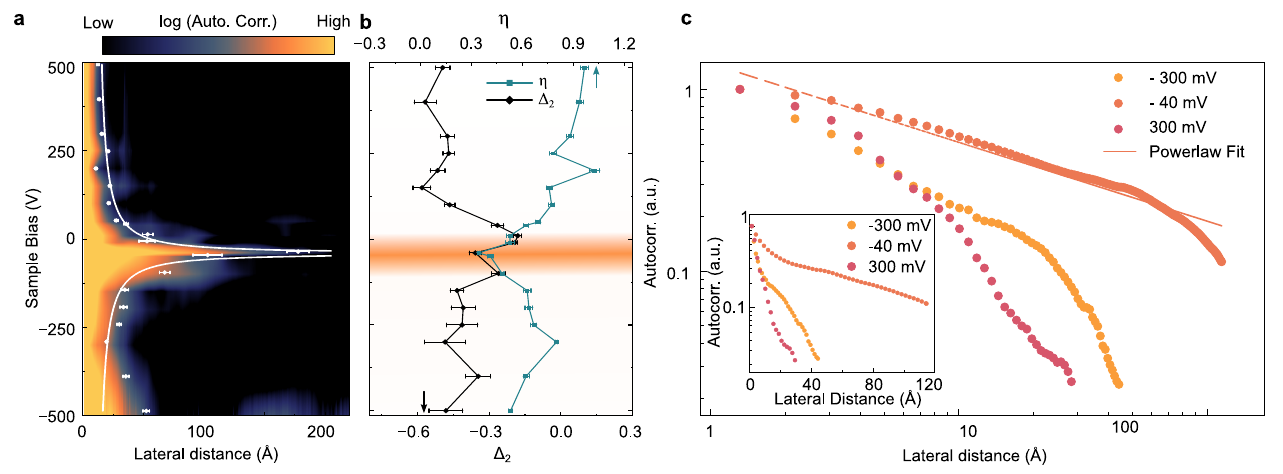}
\caption{\textbf{Correlation length divergence and power-law decay.} \textbf{a,} 2D-autocorrelation function, as a function of energy calculated from measured d$I$/d$V$ maps of MAC, showing a critical - like divergence at $E-E_F= -40$ meV. The solid white line represents energy-dependent decay of the correlation length, $\xi(E) \sim (E - E_F)^{-\nu}$.  
\textbf{b,} Comparison of the energy dependent anomalous scaling dimension (\(\Delta_2\)) and the power-law decay coefficient ($\eta$) of the spatial correlation  for  MAC showing $\eta = -\Delta_2$ relation holds only for energy close to $E=E_F$, where $\xi(E)$ diverges \textbf{c,} Auto-correlation decay at the divergence and deep within the band ($\pm 300 $ meV) following a power-law decay only near the divergence. At higher energies the decay follows approximately an exponential, indicating localized states. Inset: Autocorrelation decay plotted on a semi-logarithmic scale.    
}
\label{fig3}
\end{figure*}
This is confirmed by our analysis presented in Fig.\ \ref{fig2}i-l, showing a broader and asymmetric $f(\alpha)$ that, different from graphene, indicates the coexistence of states with widely varying scaling exponents $\alpha$. The calculated singularity spectra (Fig. \ref{fig2}j) reproduce the experimental curves well and further establish that MAC hosts critical-like multifractal wavefunctions, consistent with its nature being a disordered 2D electronic system in the vicinity of a localization \textendash\ delocalization transition.

The multifractal nature of MAC is further quantified by a so-called the anomalous scaling dimension \(\Delta_q\)~\cite{evers_anderson_2008,rodriguez_multifractal_2011} (see supplementary information for detail), which characterizes the deviation of wavefunction moments from simple dimensional scaling ($d=2$).  The anomalous scaling dimension provides a quantitative measure of multifractality in disordered systems, with a non-zero $\Delta_q$ indicating spatially fluctuating, multifractal eigenstates, whereas $\Delta_q=0$ for all $q$, for homogeneous extended states. Different from graphene, where we find $\Delta_q \approx 0$, in MAC, both experiment (Fig.\ \ref{fig2}k) and calculations (Fig.\ \ref{fig2}l) confirm that $\Delta_q$ follows a pronounced inverted parabolic dependence on $q$, confirming multifractality. A parabolic anomalous dimension $\Delta_q$ implies a parabolic $f(\alpha)$ and reflects log‑normal wave‑function statistics governed by an effectively Gaussian field theory~\cite{evers_anderson_2008}.

Anderson criticality is conventionally identified through the divergence of the electronic localization length $\xi(E)$ \cite{Krameri1993} which marks the transition between the exponentially localized and extended electronic states. Within the single-parameter scaling framework~\cite{abrahams_scaling_1979,evers_anderson_2008}, the divergence of $\xi(E)$  follows a power-law $\xi(E) \sim (E - E_c)^{-\nu}$ near the critical energy $E_c$, where the universal critical exponent $\nu$~\cite{evers_anderson_2008} is determined by the system's symmetry class.

The localization length obtained from a radial average of the two-point auto-correlation of LDOS maps at each energy is shown in Fig.\ \ref{fig3}. Fig.\ \ref{fig3}a shows the 2D autocorrelation decay, plotted as a function of energy. Uniquely, we observe a pronounced divergence of $\xi(E)$ at   $E-E_F =$ -40 meV near the band centre. This confirms that electronic states exhibit long-range spatial correlations. A power-law fit to $\xi(E)$ yields $\nu = 0.90 \pm 0.03$.
%
%

MAC can be expected to possess negligible intrinsic spin-orbit coupling and lacks global sublattice symmetry due to the random network of $sp^2$ - bonds. Consequently, MAC can not  be expected to fall into the symplectic ($\nu \approx 2.05 -2.88$~\cite{asada2002anderson}) or the ideal chiral ($\nu \approx 0.2 - 0.6$~\cite{eilmes2004exponents}) classes. 
The intermediate value of $\nu$ observed in MAC, thus likely points to unconventional criticality tied to the constrained network topology rather than any of the classified conventional Anderson universality classes~\cite{evers_anderson_2008}. 
This interpretation is further supported by the multifractal strength $\gamma$ extracted from parabolic fitting the singularity spectrum (see supplementary information). At $-40$ meV, where $\xi(E)$ diverges, we find $\gamma \approx 0.05$ closely matching that numerically obtained ($\gamma \approx 0.09$). However, the $\gamma$ differ from the numerically expected values for symplectic ($\gamma \approx 0.19$~\cite{minakuchi_two-dimensional_1998, evers_anderson_2008, schweitzer_multifractal_1995}) or unitary classes ($\gamma \approx 0.27$~\cite{schweitzer_disorder-driven_2008, potempa_localization_1999}).
It thus uniquely reflects an unconventional critical-like behavior in a truly amorphous lattice in which all spatial symmetries are broken.

A power-law decay of spatial correlations is a well-established signature of critical or near-critical electronic states at the Anderson transition, where the localization length diverges and wavefunctions exhibit multifractal spatial structure~\cite{evers_anderson_2008, cuevas_two-eigenfunction_2007}. Comparison of the auto-correlation decay near the band centre ($E-E_F= -40$ meV), with that deep within the band ($\pm 300 meV$) shows a slow, power-law decay \textit{only} near $E=E_F$ (solid line in Fig.\ \ref{fig3}c). This is distinctly different from the behavior at higher energies, where the correlation length decays rapidly.


The theoretical description of critical wavefunctions predicts that spatial correlations are linked to multifractality through the scaling relation $\eta = -\Delta_2$~\cite{evers_anderson_2008}. We extract the power-law decay exponent of the spatial correlation $\eta$ at all energies (Fig.\ \ref{fig3}b) and compared it vis-a-vis with the anomalous exponent $\Delta_2$ ($\Delta_q$ for $q=2$), obtained independently from the same LDOS map using multifractal analysis (see supplementary Methods). The comparison in Fig.\ \ref{fig3}b clearly shows that the multifractal scaling relation strictly holds \textit{only} near $E\sim0$ where the correlation length diverges. 
In contrast, for graphene, $\Delta_2$ remains approximately energy-independent and close to zero (see supplementary information), consistent with the fully extended character of its metallic electronic states.
The simultaneous observation of a diverging correlation length $\xi(E)$, the power-law decay of the auto-correlation function and multifractal scaling of the LDOS collectively suggest that monolayer amorphous carbon 
hosts critical-like extended electronic states near the band centre $E\sim0$. Beyond the known identification of critical states in disordered materials~\cite{richardella_visualizing_2010, shin_structural-disorder-driven_2023,jack_visualizing_2021}, we thus demonstrate internal consistency through independent verification of the multifractal scaling relation $\eta = -\Delta_2$~\cite{evers_anderson_2008}, which is satisfied only at the same energy where the correlation length diverges. Together with the quantitative agreement between experiment and atomistic tight-binding calculations, these observations establish the presence of a stabilized critical-like electronic state in a strictly 2D amorphous lattice, extending the experimentally validated phenomenology of Anderson criticality beyond crystalline and substitutionally disordered systems~\cite{richardella_visualizing_2010}.

\section{Discussion}\label{sec12}
The persistence of a scale-invariant, critical-like electronic state at a single energy ($E \sim 0$) in a strictly 2D system is incompatible with conventional one-parameter scaling for non-interacting electrons in the absence of additional constraints. In MAC, disorder originates not from random on-site potentials but from the topology of the continuous random network itself, preserving a predominantly off-diagonal hopping structure at low energies despite the absence of translational or exact sublattice symmetry. This distinction is crucial: while individual non-hexagonal rings locally break bipartiteness, the statistical connectivity of the $sp^2$-bonded network constrains quantum interference globally, allowing remnant chiral structure to survive in an emergent form. In the language of field theory, such constrained hopping disorder necessarily generates an additional topological contribution to the nonlinear sigma model, most naturally captured by a Wess-Zumino-Witten (WZW) topological term~\cite{wess1971consequences,witten1983global}. The role of this term is not to enforce metallic behavior, but to modify renormalization group flow such that complete localization at the band centre ($E \sim 0$) is avoided, stabilizing a scale-invariant fixed point instead. The existence of power-law spatial correlations, multifractal wavefunctions, and the selective validity of the scaling identity $\eta = -\Delta_2$ at $E \sim 0$ are direct manifestations of this modified flow. Within this framework, disclination defects such as pentagons and heptagons are described by effective gauge fields that encode the nontrivial boundary conditions generated by removing or inserting lattice wedges (see Supplementary Information for details). Importantly, these gauge fields only mix the ($A_K$,$B_{K}\prime$) and ($A_{K}\prime$,$B_K$) sectors, such that the Hamiltonian in this basis retains a purely off-diagonal structure, constituting the unexpected emergent low-energy chiral structure relevant for the observed criticality.

We note that this low-energy chiral structure should be distinguished from the chiral sublattice (A–B) symmetry retained in bipartite systems with purely off-diagonal disorder, such as graphene with random nearest-neighbour hoppings or vacancies, but evidently broken by the odd-membered rings in MAC. Consequently, this sets MAC apart from the G\"{a}de-Wegner theory~\cite{gade1991n,gade1993anderson}, which predicts a diverging density of states at the band centre, in contrast with the symmetric but suppressed density of states observed near the Fermi energy (Fig. 1d). Indeed, the effective gauge disorder associated with the topological defects, together with the emergent low-energy chiral structure discussed above, naturally suggests a Wess-Zumino-Witten (WZW) topological term in the corresponding nonlinear sigma model (see Supplementary Information for details).

Physically, this topological term modifies quantum interference between multiple scattering paths and suppresses coherent backscattering, thereby preventing the conventional Anderson-localized phase and instead stabilizing critical states at the band centre. Its predictions align with the coexistence of strong spatial inhomogeneity and extended-like behavior observed in our d$I$/d$V$ maps through multifractal wavefunctions and the absence of conventional exponential localization at E = 0. Importantly, this mechanism is not tied to translational symmetry or crystalline order, but in stead emerges from the topology and connectivity of the disordered network. As such, we highlight the excellent quantitative agreement between the nearest neighbour tight binding single-state data shown in Supplementary Fig. S1 and Fig. S2 and the theoretically predicted algebraic decay exponent $\eta$ the multifractal exponent $\gamma$, and the density-of-states scaling exponent z, as a valuable sanity check for the WZW scenario.

\section{Conclusion}\label{sec13}
In conclusion, we have shown from  multifractal analysis and spatial correlation of the local density of states that 2D monolayer amorphous carbon (MAC) supports a critical-like extended electronic state near $E\sim 0$, protected by emergent chiral symmetry. While states are exponentially localized away from the band centre, we suspect that remnant chiral structure of the $sp^2$-bonded 2D network supports a Wess \textendash\ Zumino \textendash\ Witten (WZW) topological term which modifies quantum interference and suppresses localization. 
Our findings demonstrate that scale‑invariant electronic states can be stabilized by emergent symmetry and connectivity in topologically disordered amorphous 2D networks, establishing MAC as a unique system to explore topology \textendash\ driven quantum criticality beyond the standard Anderson framework.

\section{Methods} 
\textbf{Gr/SiC sample preparation.} Epitaxial graphene substrates were obtained by flash annealing 6H-SiC(0001) substrates in an ultra-high vacuum (UHV) preparation chamber with a base pressure of $4 \times 10^{-10}$~mbar. 

\textbf{Synthesis of MAC:}
Monolayer amorphous carbon (MAC) was synthesized via laser-assisted self-limited growth, as previously reported~\cite{toh_synthesis_2020}. A $35~{\mu}m$-thick Ar/H$_2$-annealed copper (Cu) foil was loaded into the reactor chamber. Growth was initiated by applying an 80 W plasma and 248 nm laser radiation to the Cu surface for 10 minutes under 20 sccm methane (CH$_4$) at $\approx 1.5 \times 10^{-2}$~mbar. The MAC film forms on the backside of the Cu foil.

Subsequently, the MAC film was transferred to a flash-annealed Au on Mica substrate using a polymer-free method to minimize organic contamination. First, to ensure only the MAC layer (formed on the back) was analyzed, the undesirable top-surface carbon film was removed using an oxygen plasma etch prior to transfer. The remaining MAC/Cu foil was then placed on a 0.7 $\%$  Ammonium Persulfate (APS) solution and left overnight until the Cu fully dissolved. The resulting free-floating MAC film was rinsed in a beaker of deionized (DI) water for 3 hours before being scooped onto the target substrate. The final sample was air-dried for at least 2 hours prior to transfer into the scanning tunneling microscope (STM) storage chamber.

The structure of MAC has been verified using atomic resolution transmission electron microscopy, Raman spectroscopy, and X-ray photoemission spectroscopy to ensure competing nanocrystalline continious random network atomic organisation in agreement with earlier reports.

\textbf{Scanning tunnelling spectroscopy:} Low-temperature scanning tunnelling microscopy and spectroscopy (STM/STS) was performed in an Scienta Omicron STM at $T\simeq 4.5$~K under UHV conditions ($\approx 5 \times 10^{-11}$~mbar). In all STM measurements, we used an electrochemically etched tungsten tip calibrated against the Au(111) Shockley surface state. Spectroscopy measurements were performed using standard lock-in techniques with a modulation amplitude of $V_{\rm{mod}}$= 20~mV and a modulation of frequency of 732 Hz unless otherwise specified. Differential conductance maps were taken in constant current mode.

\textbf{Numerical methods:}
Numerical calculations were carried out on $45\,\mathrm{nm} \times 45\,\mathrm{nm}$ MAC structures obtained using a two-dimensional restricted quench-from-the-melt procedure following~\cite{zhang_superior_2025}. The $45\,\mathrm{nm} \times 45\,\mathrm{nm}$ simulation cell contains $75{,}000$ carbon atoms with a hexagonal ring content of $74\%$ obtained by employing a cooling rate of $0.1\,\mathrm{K\,fs^{-1}}$. A bond-length-independent nearest-neighbour hopping model was employed to isolate the effects of topological disorder in MAC. Electronic structure calculations were performed using a nearest-neighbour tight-binding model with a bond-length-independent hopping parameter $t = -2.7\,\mathrm{eV}$. The amorphicity-dependent DOS (Fig 1.g)  was constructed with smaller $24\,\mathrm{nm} \times 24\,\mathrm{nm}$ MAC structures, and the hexagonal ring content was tuned ($71-89\%$) by adjusting the cooling rate. Averages were taking over 2 independent MAC samples per amorphicity. For the result presented in \ref{fig2} the LDOS was broadened by $20\,\mathrm{meV}$ to account for the experimental lock-in modulation amplitude. Histograms and multifractal analyses were carried out using box sizes of $0.35\,\mathrm{nm}$ and $0.7\,\mathrm{nm}$, corresponding to the experimental parameters used.
Unbroadened LDOS was used to obtain the correlation presented in Fig S1, the multifractal analysis in S1. All data were averaged over 9 different MAC samples. 


\section{Acknowledgement} 
This work was supported by Singapore National Research Foundation-Frontier of Competitive Research Programme (NRF-F-CRP-2024-0012) and partialsupport from the Singapore Ministry of Education (MOE) Academic Research Fund Tier 3 grant (MOE-MOET32023-0003) ‘Quantum Geometric Advantage’.


\section{Author Contributions} 
RSK performed the scanning tunnelling microscopy and spectroscopy experiments with help from RS  GS and ZJT. AKG CTT and HZ prepared the MAC samples. RSK and BW analyzed the STM/STS data. RSK performed the multifractal analysis with the help of BW, RS and AN. NK, YS and KS performed the TEM experiment. HZ and OVY performed the tight-binding calculation. AH and HZ conceptualized the study and carried out the field-theoretical calculations related to the WZW physics. SA, RAR and OVY supervised the theory work.
BW  and BO conceived the idea and coordinated the project.
RSK and BW wrote the manuscript with input from all authors

\section{Competing Interests}
The authors declare no competing interests.


\bibliography{reference.bib}

\end{document}